\documentclass[aps,prd,preprint,superscriptaddress,showpacs]{revtex4}
\usepackage{graphicx}

\def\bwt{\begin{widetext}}
\def\ewt{\end{widetext}}
\def\be{\begin{equation}}
\def\ee{\end{equation}}
\def\bea{\begin{eqnarray}}
\def\eea{\end{eqnarray}}
\def\bean{\begin{eqnarray*}}
\def\eean{\end{eqnarray*}}
\def\bary{\begin{array}}
\def\eary{\end{array}}
\def\bit{\begin{itemize}}
\def\eit{\end{itemize}}

\def\su5u1{SU(5) \times U(1)}
\def\fsu5u1{SU(5) \times U(1)'}
\def\so10{SO(10)}
\def\sq20{SO(10) \times SO(10)}

\usepackage[centertags]{amsmath}
\usepackage{amssymb}
\newcommand{\Z}{{\mathbb Z}}

\begin{document}

\title{Standard-Like Model Building on Type II Orientifolds}

\author{Ching-Ming Chen}

\affiliation{George P. and Cynthia W. Mitchell Institute for
Fundamental Physics,
 Texas A$\&$M University, College Station, TX 77843, USA }

\author{Tianjun Li}

\affiliation{Department of Physics and Astronomy, Rutgers University, 
Piscataway, NJ 08854, USA }

\affiliation{Institute of Theoretical Physics, Chinese Academy of Sciences,
 Beijing 100080, P. R. China }

 \author{Dimitri V. Nanopoulos}

\affiliation{George P. and Cynthia W. Mitchell Institute for
Fundamental Physics,
 Texas A$\&$M University, College Station, TX 77843, USA }

\affiliation{Astroparticle Physics Group, 
Houston Advanced Research Center (HARC),
Mitchell Campus, Woodlands, TX 77381, USA}

\affiliation{Academy of Athens, Division of Natural Sciences, 
 28 Panepistimiou Avenue, Athens 10679, Greece }

\date{\today}

\begin{abstract}
We construct new Standard-like models on Type II orientifolds.
In Type IIA theory on $\mathbf{T^6/(\Z_2\times \Z_2)}$  orientifold 
with intersecting D6-branes, we first construct a three-family 
trinification model where the $U(3)_C\times U(3)_L\times U(3)_R$ 
gauge symmetry can be broken down to the 
$SU(3)_C\times SU(2)_L\times U(1)_{Y_L}\times U(1)_{I_{3R}}\times U(1)_{Y_R}$ 
gauge symmetry by the Green-Schwarz mechanism and D6-brane splittings, 
and further down to the SM gauge symmetry  at the TeV scale 
by Higgs mechanism. We also construct a Pati-Salam  model
where  we may explain three-family SM fermion masses and mixings.
Furthermore, we construct for the first time a Pati-Salam like model with 
$U(4)_C \times U(2)_L \times U(1)' \times U(1)''$ gauge symmetry
where the $U(1)_{I_{3R}}$  comes from a linear combination of 
$U(1)$ gauge symmetries. In Type IIB theory on $\mathbf{T^6/(\Z_2\times \Z_2)}$  
orientifold with flux compactifications, we construct a new flux model with 
$U(4)_C \times U(2)_L \times U(2)_R$ gauge symmetry where the magnetized 
D9-branes with large negative D3-brane charges are introduced in 
the hidden sector. However, we can not construct the trinification
model with supergravity fluxes because the three $SU(3)$ groups
already contribute very large RR charges. The phenomenological 
consequences of these models are briefly discussed as well.

\end{abstract}


\preprint{ACT-08-05, MIFP-05-21, hep-th/0509059}

\maketitle

\section{Introduction}

The major goal of string phenomenology is to construct 
four-dimensional models with the features of
the Standard Model (SM), which connect the string theory to 
realistic particle physics. 
Due to the advent of D-branes~\cite{JPEW}, we can construct 
consistent four-dimensional chiral models with
non-Abelian gauge symmetry on Type II orientifolds. 
Chiral matter can appear: (i) due to the D-branes located at orbifold
singularities with chiral fermions appearing on the worldvolume of
such D-branes~\cite{ABPSS,berkooz}; (ii) at the intersections of D-branes
in the internal space~\cite{bdl} with T-dual description in terms 
of magnetized D-branes~\cite{bachas}.

On Type IIA orientifolds with intersecting D6-branes, 
 a large number of non-supersymmetric three-family
Standard-like models and grand unified models, which
satisfy the Ramond-Ramond (RR) tadpole
cancellation conditions, were 
constructed~\cite{LU,IB,IBII,LUII,Sagnottietal,CIM,List}. 
However, there generically exist two problems: the
uncancelled Neveu-Schwarz-Neveu-Schwarz (NSNS) tadpoles
and the gauge hierarchy problem.
On the other hand,
the first supersymmetric models~\cite{CSU1,CSU2} with
quasi-realistic features of the supersymmetric Standard-like
models have been constructed in Type IIA theory on
$\mathbf{T^6/(\Z_2\times \Z_2)}$ orientifold with intersecting D6-branes.
Then,  supersymmetric Standard-like models,
 Pati-Salam like models, $SU(5)$ models as well as 
 flipped $SU(5)$ models have been constructed 
systematically~\cite{CP,Cvetic:2002pj,CLL,Cvetic:2004nk,Chen:2005ab}, and 
their  phenomenological consequences have been
studied~\cite{CLS1,CLS2,CLW}.
In addition, supersymmetric constructions on other Type IIA 
orientifolds were also discussed~\cite{ListSUSYOthers}.

In spite of these successes, the moduli stabilization in open string and
closed string sectors is still an open problem even if  some of the 
complex structure parameters (in  the Type IIA picture)  
and dilaton fields may be
stabilized due to the gaugino
condensation in the hidden sector in some models (see, e.g., \cite{CLW}.).
The supergravity fluxes give us another way to stabilize the 
compactification moduli 
fields by lifting continuous moduli space of the string vacua in an effective
 four-dimensional theory (see, e.g., \cite{GVW})
because it introduces a supergravity potential. 
In this paper, for flux model building, we only consider the
Type IIB orientifolds.
The intersecting D6-brane constructions correspond to the models with
magnetized D-branes where the role of the  intersecting angles is
played by the magnetic fluxes on the D-branes on the Type IIB orientifolds.
In~\cite{CU,BLT}, the techniques  for consistent chiral 
flux compactifications on Type IIB 
 orientifolds were developed, and the dictionary for  the consistency and 
supersymmetry conditions between the two T-dual constructions was given.
 However, due to the Dirac quantization conditions, 
the supergravity fluxes impose strong constraints on 
consistent model building, 
since they contribute large positive D3-brane charges and then
 modify the global RR tadpole cancellation conditions significantly.
Therefore, no explicit supersymmetric chiral Standard-like models
were obtained in~\cite{CU,BLT}.
In Type IIA orientifolds with flux compactifications, 
it has been recently shown that
the RR, NSNS and metric fluxes could contribute
negative D6-brane charges, and then relax the RR tadpole 
cancellation conditions~\cite{Villadoro:2005cu,Camara:2005dc}.
But we will not consider it in this paper.

 By introducing the magnetized D9-branes with large negative D3-brane
charges in the hidden sector,   the first three-family and four-family 
Standard-like models with one unit of quantized flux were obtained~\cite{MS,CL}.
  These constructions are T-dual to the supersymmetric models of 
intersecting D6-branes on $\Z_2\times \Z_2$ 
orientifold with the $Sp(2)_L\times Sp(2)_R$ or $Sp(2f)_L\times Sp(2f)_R$
gauge symmetry in the electroweak sector, 
respectively~\cite{Cvetic:2004nk}
(For the corresponding non-supersymmetric models, see~\cite{CIM}.). 
Recently this kind of models has been studied systematically~\cite{Kumar:2005hf}. 
Moreover, considering the magnetized D9-branes with large negative D3-brane 
charges in the SM observable sector, a lot of new models have been 
constructed~\cite{Cvetic:2005bn}, for example, 
many new models with one unit of quantized flux, 
the first three- and four-family models with two units of quantized fluxes,
and the first three- and four-family models with supersymmetric flux, 
{\it i.~e.}, three units of quantized fluxes.

However, in the previous model building with or without fluxes,
one of the serious problems is how to generate 
 suitable three-family SM fermion masses
and mixings. In the $SU(5)$ models~\cite{Cvetic:2002pj} 
and the flipped $SU(5)$ models~\cite{Chen:2005ab}, the up-type
quark Yukawa couplings and the down-type quark Yukawa couplings are 
forbidden by 
anomalous $U(1)$ gauge symmetries. And in the Pati-Salam like models,
although we can have the Yukawa couplings in principle, 
it is very difficult to construct three-family models which can give 
 suitable SM fermion masses and mixings for three families
because the left-handed fermions,
the right-handed fermions and the Higgs fields in general arise from 
the intersections on different two-tori~\cite{CLLL-P}. 
 In addition, if supersymmetry is broken by supergravity fluxes, 
it seems that the masses for the rest massless SM fermions 
may not be generated from radiative corrections~\cite{CLLL-P}
 because the supersymmetry breaking trilinear soft terms are
universal and the supersymmetry breaking soft masses for the left/right-chiral
squarks and sleptons are also 
universal~\cite{Soft BreakingI,Soft BreakingII,Soft BreakingIII} 
since the K\"ahler potential for
the SM fermions depends only on the intersection 
angles~\cite{CPY,LustetalIII,Abel:2004ue}.
Thus, how to  generate suitable three-family SM fermion
masses and mixings is an interesting question.
Another interesting question is whether we can constuct new models with
and without fluxes.

In this paper,  in Type IIA theory on
$\mathbf{T^6/(\Z_2\times \Z_2)}$ orientifold with intersecting D6-branes,
we first construct a three-family
trinification model~\cite{RGG,Babu:1985gi} 
with initial $U(3)_C \times U(3)_L \times U(3)_R$ gauge symmetry,
which have not been studied previously (For the three-family trinification 
model building in the $Z_3$ orbifold compactification of
weakly coupled heterotic string theory, see, e.g.,~\cite{Choi:2003ag}). 
Because of large RR charges from three $U(3)$ groups,
it is very difficult to satisfy the RR tadpole cancellation conditions.
In the trinification models, all
the SM fermions and Higgs fields belong to the bi-fundamental
representations, which can naturally arise from the intersecting
D6-brane model building. Especially, the bi-fundamental representation
with one fundamental and one anti-fundamental indices is
different from the  bi-fundamental representation 
with two  fundamental (or anti-fundamental) indices, for example, 
$(\mathbf{3, \overline{3}, 1})$ and $(\mathbf{3, 3, 1})$ 
(or $(\mathbf{\overline{3}, \overline{3}, 1})$).
So,  the three-family trinification models can be 
constructed even if there is no tilted two-torus. Moreover,
the $U(3)_C\times U(3)_L\times U(3)_R$ gauge symmetry is
broken down to the $SU(3)_C\times SU(3)_L\times SU(3)_R$ gauge symmetry
due to the Green-Schwarz (G-S) mechanism. Also,
the $SU(3)_C\times SU(3)_L\times SU(3)_R$ gauge symmetry can be
broken down to the $SU(3)_C\times SU(2)_L\times U(1)_{Y_L}\times
U(1)_{I_{3R}}\times U(1)_{Y_R}$ gauge symmetry by the splittings of
the $U(3)_L$ and $U(3)_R$ stacks of 
the D6-branes~\cite{Cvetic:2004nk,Choi:2005pk}, and the 
$U(1)_{Y_L}\times U(1)_{I_{3R}}\times U(1)_{Y_R}$ gauge symmetry can
be broken down to the $U(1)_Y$ gauge symmetry by the Higgs mechanism 
at the TeV scale which can not preserve the D-flatness and F-flatness 
and then breaks four-dimensional $N=1$ supersymmetry
 because we assume the low scale supersymmetry 
in this paper. In addition, 
the quark Yukawa couplings are allowed while the
lepton  Yukawa couplings are forbidden by 
anomalous $U(1)$ gauge symmetries.
In our model, we can only give masses to one
family of the SM quarks.

For the Pati-Salam model, we consider
the particles with quantum numbers $(\mathbf{4, \overline{2}, 1})$
and $(\mathbf{ \overline{4}, 1, 2})$  
under the $U(4)_C \times U(2)_L \times U(2)_R$ gauge symmetry as
the SM fermions while we consider the
particles with quantum numbers $(\mathbf{ 4, 2, 1})$ 
and $(\mathbf{ \overline{4}, 1, \overline{2}})$ as exotic particles
because these particles are in fact 
distinguished by anomalous $U(1)$ gauge 
symmetries (In the trinification models, these kinds of particles
are obviously different.). With this convention,
we construct a three-family Pati-Salam model without tilted two-torus 
where the left-handed and right-handed
SM fermions and the Higgs fields arise from the intersections on the
same two-torus, and then, we may explain three-family 
SM fermion masses and mixings. Also,
the $U(4)_C \times U(2)_L \times U(2)_R$ gauge symmetry
can be broken down to the 
$SU(3)_C \times SU(2)_L \times U(1)_{I_{3R}} \times U(1)_{B-L}$ 
gauge symmetry
due to the G-S mechanism and D6-brane splittings. 
In our  model, the $U(1)_{I_{3R}} \times U(1)_{B-L}$
gauge symmetry can only be broken down to the $U(1)_Y$ gauge
symmetry at the TeV scale by Higgs mechanism.

In all the previous Pati-Salam like model building,  the
$U(1)_{I_{3R}}$ gauge symmetry arises from the non-Abelian 
gauge symmetry, for example, $U(2)_R$ or $USp(2f)_R$.
 We first construct a Pati-Salam like model with  
 $U(4)_C \times U(2)_L \times U(1)' \times U(1)''$ 
gauge symmetry where the 
$U(1)_{I_{3R}}$  comes from a linear combination of
  $U(1)$ gauge symmetries. Also, 
 the $U(1)_{I_{3R}} \times U(1)_{B-L}$
gauge symmetry can be broken down to the $U(1)_Y$ gauge
symmetry at  (or close to) the string scale by Higgs mechanism.  
However, only one family of the SM fermions can obtain masses.

In Type IIB theory on $\mathbf{T^6/(\Z_2\times \Z_2)}$
 orientifold with flux compactifications, 
 we point out that one may not construct the trinification
model with supergravity fluxes because the three $SU(3)$ groups
already contribute very large RR charges. In addition,
we construct a new flux model with 
$U(4)_C \times U(2)_L \times U(2)_R$ gauge symmetry where the
 magnetized D9-branes with large negative D3-brane
charges are introduced in the hidden sector. This kind of models
has not been studied previously because it is very difficult to
have supersymmetric D-brane configurations 
with more than three stacks
of $U(n)$ branes. However, in our model,
 the $U(1)_{I_{3R}} \times U(1)_{B-L}$
gauge symmetry can only be broken down to the $U(1)_Y$ gauge
symmetry at the TeV scale by Higgs mechanism, and we can only give 
masses to one family of the SM fermions.

This paper is organized as follows. In Section II, 
we  review  the D-brane model building in
Type II theories on the $\mathbf{T^6/(\Z_2\times \Z_2)}$  orientifolds.
We consider the Type IIA non-flux model building and the 
Type IIB flux model building in Sections III and IV, respectively.
Section V is our discussions and conclusions.

\section{Intersecting D-Brane Model Building}

In this Section, we briefly review the rules for the
intersecting D-brane model building in Type II 
theories on $\mathbf{T^6/(\Z_2\times \Z_2)}$ 
orientifolds~\cite{CSU1,CSU2,CU,BLT}.
It is well known that a supersymmetric Type IIA intersecting D6-brane
construction is T-dual to a Type IIB construction
 with magnetized D3-, D5-, D7-, and D9-branes, thus,
the rules for their model building are quite similar.
For Type IIB constructions, we shall consider  
additional supergravity fluxes which
give us stronger constraint on the RR tadpole cancellation conditions.

\subsection{Type IIA Construction}

We first briefly review the intersecting D6-brane model 
building in Type IIA theory on
$\mathbf{T^6/(\Z_2\times \Z_2)}$  orientifold~\cite{CSU1,CSU2}.
 We consider $\mathbf{T^{6}}$ to be a
six-torus factorized as 
$\mathbf{T^{6}} = {\bf T}^{2} \times {\bf T}^{2} \times {\bf T}^{2}$
whose complex coordinates are $z_i$, $i=1,\; 2,\; 3$ for the
$i$-th two-torus, respectively. The $\theta$ and $\omega$
generators for the orbifold group $\Z_{2} \times \Z_{2}$
 act on the complex coordinates as following 
\begin{eqnarray}
& \theta: & (z_1,z_2,z_3) \to (-z_1,-z_2,z_3)~,~ \nonumber \\
& \omega: & (z_1,z_2,z_3) \to (z_1,-z_2,-z_3)~.~\,
\label{Z2Z2} 
\end{eqnarray}
We implement an orientifold projection $\Omega R$, where $\Omega$
is the world-sheet parity, and $R$ acts on the complex coordinates as
\begin{equation}
R:(z_1,z_2,z_3)\rightarrow(\overline{z}_1,\overline{z}_2,\overline{z}_3)~.~\,
\end{equation}

With the wrapping numbers $(n^i, m^i)$ along the canonical bases
of homology one-cycles $[a_i]$ and $[b_i]$, the general homology class
for one cycle on the $i$-th two-torus
${\bf T}^2_i$ is given by $n^i[a_i]+m^i[b_i]$. Therefore, the
complete homology classes for the three cycles wrapped by a stack of $N_a$
D6-branes $[\Pi_a]$ and its orientifold image $[\Pi_a']$ can be written as
\begin{equation}
[\Pi_a]=\prod_{i=1}^{3}(n^i_a[a_i]+m^i_a[b_i])~,~\;\;\;
[\Pi_{a'}]=\prod_{i=1}^{3}(n^i_a[a_i]-m^i_a[b_i])~.~\,
\end{equation}

The  homology classes for
four O6-planes associated with the orientifold projections
$\Omega R$, $\Omega R\theta$, $\Omega R \omega$, and $\Omega
R\theta\omega$  are
\begin{eqnarray}
\Omega R :&[\Pi^{(1)}]=[a_1][a_2][a_3] ~,~\nonumber \\
\Omega R \omega :&[\Pi^{(2)}]=-[a_1][b_2][b_3] ~,~\nonumber \\
\Omega R \theta \omega :&[\Pi^{(3)}]=-[b_1][a_2][b_3] ~,~\nonumber \\
\Omega R \theta :&[\Pi^{(4)}]=-[b_1][b_2][a_3]  ~.~\label{O6IIA}
\end{eqnarray}

 The four-dimensional $N=1$
supersymmetric models from Type IIA orientifolds with intersecting
D6-branes are mainly constrained in two aspects: RR tadpole
cancellation conditions  and four-dimensional $N=1$ supersymmetry conditions:

(1) RR Tadpole Cancellation Conditions

 The total RR charges of D6-branes and O6-planes must vanish since
the RR field flux lines are conserved. With the filler branes on the
top of the four O6-planes, we have the RR tadpole cancellation conditions:
\begin{eqnarray}
&& -N^{(1)}-\sum_a N_a n^1_a n^2_a n^3_a = 
-N^{(2)}+\sum_a N_a n^1_a m^2_a m^3_a= \nonumber \\
&& -N^{(3)}+\sum_a N_a m^1_a n^2_a m^3_a = -N^{(4)}+\sum_a N_a
m^1_a m^2_a n^3_a = -16 ~,~ \label{RRIIA}
\end{eqnarray}
where $N^{(i)}$ denotes the number of the filler branes on the top of the
$i$-th O6-brane defined in Eq. (\ref{O6IIA}).

(2) Four-Dimensional $N=1$  Supersymmetry Conditions

The four-dimensional $N=1$ supersymmetry  can be preserved by the
orientation projection if and only if the rotation angle of any
D6-brane with respect to the orientifold-plane is an element of
$SU(3)$~\cite{bdl}, {\it i.~e.},
$\theta_1+\theta_2+\theta_3=0 $ mod $2\pi$, where $\theta_i$ is
the angle between the $D6$-brane and the orientifold-plane on the
$i$-th two-torus. This supersymmetry condition can be
rewritten as~\cite{Cvetic:2002pj}
\begin{eqnarray}
-x_A m^1_a m^2_a m^3_a + x_B m^1_a n^2_a n^3_a + x_C n^1_a m^2_a
n^3_a  + x_D n^1_a n^2_a m^3_a =0~,~
\nonumber \\
-n^1_a n^2_a n^3_a/x_A + n^1_a m^2_a m^3_a/x_B + m^1_a n^2_a
m^3_a/x_C + m^1_a m^2_a n^3_a/x_D <0 ~,~\,
\label{SUSYIIA}
\end{eqnarray}
where $x_A=\lambda,\;
x_B=\lambda 2^{\beta_2+\beta3}/\chi_2\chi_3,\; x_C=\lambda
2^{\beta_1+\beta3}/\chi_1\chi_3,\; x_D=\lambda
2^{\beta_1+\beta2}/\chi_1\chi_2$,  $\chi_i=R^2_i/R^1_i$ are the
complex structure parameters and $\lambda$ is a positive real number.

\subsection{Type IIB Construction}

We consider the Type IIB flux compactifications on the same 
orientifold $\mathbf{T^6/(\Z_2\times \Z_2)}$~\cite{CU,BLT}.  The associated
orientifold projection is still $\Omega R$, where the corresponding 
R acts on the complex coordinates as~\cite{CU}
\begin{equation}
R:(z_1,z_2,z_3)\rightarrow(-z_1,-z_2,-z_3)~.~\,
\end{equation}

We need $D(3+2n)$-branes to fill up the four-dimentional Minkowski
space-time and wrap the $2n$-cycles on a compact manifold in
Type IIB theory.  And the introduction of magnetic fluxes provides the
consistency to Type IIB theory.  For one stack
of $N_a$ D-branes wrapping $m_a^i$ times on the $i$-{th} two-torus ${\bf T}^2_i$, we
turn on $n_a^i$ units of magnetic fluxes $F_a$ for the center of mass $U(1)_a$
gauge factor on each ${\bf T}^2_i$
\begin{eqnarray}
m_a^i \, \frac 1{2\pi}\, \int_{{\bf T}^2_{\,i}} F_a^i \, = \, n_a^i ~.~\,
\label{monopole}
\end{eqnarray}
Therefore, the
D9-, D7-, D5- and D3-branes  contain 0, 1, 2 and 3 vanishing $m_a^i$s,
respectively. The even homology classes for the point and the two-torus
on the $i$-th two-torus ${\bf T}^2_i$ are denoted as
$[{\bf 0}_i]$ and $[{\bf T}^2_i]$, respectively. Then,
the vectors of RR charges of the $a$-th stack of D-branes
 and its $\Omega R$ image are
\begin{eqnarray}
[{ \Pi}_a]\, =\, \prod_{i=1}^3\, ( n_a^i [{\bf 0}_i] + m_a^i [{\bf T}^2_i] )~,~  [{
\Pi}_a']\, =\, \prod_{i=1}^3\, ( n_a^i [{\bf 0}_i]- m_a^i [{\bf T}^2_i] )~,~
\label{homology class for D-branes}
\end{eqnarray}
respectively. Similarly, the vectors of RR charges for 
 the $O3$- and $O7_i$-planes, which respectively correspond to 
$\Omega R$, $\Omega R\omega$, $\Omega
R\theta\omega$ and $\Omega R\theta$ O-planes, are
\begin{eqnarray}
&\Omega R:&[\Pi_{O3}]= [{\bf 0}_1]\times [{\bf 0}_2]\times [{\bf 0}_3]~;~ \nonumber \\
&\Omega R\omega:&[\Pi_{O7_1}]=-[{\bf 0}_1]\times [{\bf T}^2_2]\times
[{\bf T}^2_3]~;~ \nonumber \\
&\Omega R\theta\omega:&[\Pi_{O7_2}]=-[{\bf T}^2_1]\times 
[{\bf 0}_2]\times [{\bf T}^2_3]~;~ \nonumber\\
&\Omega R\theta:&[\Pi_{O7_3}]=-[{\bf T}^2_1]\times [{\bf T}^2_2]\times [{\bf 0}_3]~.~\,
\label{O-planes}
\end{eqnarray}

It is convenient to define the RR charges carried by the magnetized
D-branes in Type IIB theory~\cite{Cvetic:2005bn}.  For one stack of
$N_a$ D-brane with wrapping numbers $(n^i_a, m^i_a)$, the RR
charges of D3-, D5-, D7-, and D9-branes are
\begin{eqnarray}
Q3_a= N_a n^1_a n^2_a n^3_a~,~ \;\; Q5_{ia} = N_a m^i_a n^j_a n^k_a~,~\nonumber\\
Q7_{ia} = N_a n^i_a m^j_a m^k_a~,~ \;\; Q9_a= N_a m^1_a m^2_a m^3_a~,~\,
\label{RRcharges}
\end{eqnarray}
where $i \not= j \not= k$.

(1) RR Tadpole Concellation Conditions

The Type IIA RR tadpole concellation conditions in 
Eq. (\ref{RRIIA}) can also be
applied exactly in Type IIB picture, except that 
there are flux contributions in Type IIB flux compactifications. 
With the filler branes on the top of  the $O3$- and $O7_i$-planes,
we obtain the RR tadpole cancellation conditions
\begin{eqnarray}
&& -N^{(O3)} - \sum_a Q3_a - \frac{1}{2}N_{flux} =
-N^{(O7_1)} + \sum_a Q7_{1a} = \nonumber \\
&& -N^{(O7_2)} + \sum_a Q7_{2a} = -N^{(O7_3)} + \sum_a Q7_{3a} =
-16~,~  \label{RRIIB}
\end{eqnarray}
where $N_{flux}$ is the amount of the fluxes turned on and quantized in
units of elementary flux which is  64 as discussed later.

(2) Four-Dimensional $N=1$ Supersymmetry Conditions

Four-Dimensional $N=1$ supersymmetric vacua from flux compactifications
require 1/4 supercharges of the ten-dimentional Type I theory be
preserved in both open and closed string sectors.
 For the closed string sector, the specific
Type IIB flux solution on orientifolds comprises of self-dual
three-form field strength~\cite{Giddings0105097,Kachru0201028}.
The D3-brane RR charges contributed from the three-form flux
$G_3=F_3-\tau H_3$ are given by
\begin{equation}
N_{flux}=\frac{1}{(4\pi^2 \alpha')^2}\int_{X_6} H_3 \wedge F_3 =
\frac{1}{(4\pi^2 \alpha')^2}\frac{i}{2\mathrm{Im}(\tau)}\int_{X_6}
G_3 \wedge \bar{G_3}~,~\,
\end{equation}
where $F_3$ and $H_3$ are respectively the 
RR and NSNS three-form field strengths, and $\tau=a+i/g_s$
is the Type IIB axion-dilaton coupling.  Dirac quantization
conditions for $F_3$ and $H_3$ on $\mathbf{T^6} /(\Z_2 \times
\Z_2)$ orientifold require $N_{flux}$ to be a multiple of 64, and the BPS-like
self-dual condition $\ast_6 G_3=iG_3$ demands $N_{flux}$ to be
positive.  Supersymmetric configuration
implies that $G_3$ background field should be
a primitive self-dual (2,1) form, and a specific supersymmetric
solution is~\cite{Kachru0201028}
\begin{equation}
G_3= \frac{8}{\sqrt{3}} e^{-\pi i/6}(d\bar{z}_1 dz_2 dz_3+dz_1
d\bar{z}_2 dz_3+dz_1 dz_2 d\bar{z}_3)~.~\,
\end{equation}
This fluxes stabilize the complex structure toroidal moduli at
values
\begin{equation}
\tau_1=\tau_2=\tau_3=\tau=e^{2\pi i/3}~,~\,
\end{equation}
and its RR tadpole contribution $N_{flux}$ is 192.

In the open-string sector, 
  the four-dimensional $N=1$ supersymmetry for a D-brane configuration
is conserved  if
and only if  $\sum_i\, \theta_i=0 $ (mod $2\pi$) is satisfied~\cite{CU}
where the ``angle'' $\theta_i$  is determined in terms of the
worldvolume magnetic field  as $\tan (\theta_i)\equiv
(F^i)^{-1}=\frac{m^i\chi'^i}{n^i}$ and $\chi'^i=R^i_1R^i_2$ is
the area of the $i$-th two-torus ${\bf T}^2_i$ in $\alpha'$ units.
This condition can be rewritten in terms of RR charges as~\cite{Cvetic:2005bn}
\begin{eqnarray}
-x_A Q9_a + x_B Q5_{1a} + x_C Q5_{2a} + x_D Q5_{3a} = 0 ~,~\nonumber \\
- Q3_a /x_A + Q7_{1a} /x_B + Q7_{2a} /x_C + Q7_{3a} /x_D < 0 ~,~\,
\label{SUSYIIB}
\end{eqnarray}
where $x_A=\lambda$, $x_B=\lambda / \chi'^2\chi'^3$, $x_C=\lambda
/ \chi'^1\chi'^3$, $x_D=\lambda / \chi'^1\chi'^2$.

It is not a surprise that the four-dimensional $N=1$ supersymmetry conditions
on Type IIB orientifold are similar to those on Type IIA orientifold
in Eq. (\ref{SUSYIIA}), except the different definitions of the
complex structure parameters $\chi^i$ and $\chi'^i$.  
Therefore, we will take use of these two
kinds of conditions on equal foot when we search the models.

\subsection{Spectra}

The spectra from the Type IIA and Type IIB orientifolds are the same.
Massless chiral fields arise from the open strings with two ends
attaching on the intersections of any two different D-brane stacks. 
 The multiplicity (${\cal M}$) of
the corresponding bi-fundamental representation is given by the
intersection numbers between these two stacks of D-branes which
is determined by the wedge product of their homology classes.  
The initial $U(N_a)$ gauge group supported by
a stack of $N_a$ identical D-branes is broken down by the
$\Z_2\times \Z_2$ symmetry to a subgroup $U(N_a/2)$~\cite{CSU1,CSU2}. 
The general spectra for the D-brane models on Type II orientifolds
are given in Table \ref{Spectrum}.

\begin{table}[h]
\renewcommand{\arraystretch}{1.5}
\center
\begin{tabular}{|c||c|}
\hline

Sector & Representation   \\ \hline \hline

$aa$ & $U(N_a /2)$ vector multiplet and 3 adjoint chiral multiplets \\
\hline

$ab+ba$ & $ {\cal M}(\frac{N_a}{2},
\frac{\overline{N_b}}{2})=I_{ab}=\prod_{i=1}^3(n^i_a m^i_b - n^i_b
m^i_a) $  \\ \hline

$ab'+b'a$ & $ {\cal M}(\frac{N_a}{2},
\frac{N_b}{2})=I_{ab'}=-\prod_{i=1}^3(n^i_a m^i_b + n^i_b m^i_a) $
\\ \hline

$aa'+a'a$ & $  {\cal M}({\rm
Anti}_a)=\frac{1}{2}(I_{aa'}+\frac{1}{2}I_{aO}) $   \\
 & $ {\cal M}({\rm
Sym}_a)=\frac{1}{2}(I_{aa'}-\frac{1}{2}I_{aO})  $   \\
\hline

\end{tabular}
\caption{The general spectra for the D-brane models on Type II orientifolds, 
where $I_{aa'}=-8\prod^3_{i=1}n^i_a m^i_a$, and $I_{aO}=8(-m^1_a m^2_a
m^3_a + m^1_a n^2_a n^3_a +n^1_a m^2_a n^3_a + n^1_a n^2_a
m^3_a)$.}  \label{Spectrum}
\end{table}

A model may contain additional non-chiral (vector-like) multiplet
pairs from $ab+ba$, $ab'+b'a$, and $aa'+a'a$ sectors if two stacks of the 
corresponding D-branes are parallel and on the top of each other 
on at least one two-torus. The multiplicity of the
non-chiral multiplet pairs is given by the product of the rest
intersection numbers, neglecting the null sector. For example, if only
$(n^1_a l^1_b - n^1_b l^1_a)=0$ in $
I_{ab}=[\Pi_a][\Pi_b]=\prod_{i=1}^3(n^i_a l^i_b - n^i_b l^i_a) $, then 
we have 
\begin{equation}
{\cal M}\left[\left(\frac{N_a}{2},\frac{\overline{N_b}}{2}\right)
+\left(\frac{\overline{N_a}}{2},\frac{N_b}{2}\right)\right]
=\prod_{i=2}^3(n^i_a l^i_b - n^i_b l^i_a)~.~\,
\end{equation}

Moreover, the fermionic components of the non-chiral multiplets
may acquire tree-level masses which dependent on the 
compactification radii and the brane wrapping numbers
via Scherk-Schwarz mechanism~\cite{Angelantonj:2005hs}.

\subsection{The K-theory Conditions}

In addition to the RR tadpole cancellation conditions, the discrete D-brane RR
charges classified by $\mathbf{\Z_2}$ K-theory groups in the
presence of orientifolds, which are subtle and invisible by the
ordinary homology~\cite{MS,Witten9810188},
should also be taken into account~\cite{CU}.

In Type I superstring theory there exist non-BPS D-branes carrying
non-trivial K-theory $\mathbf{\Z_2}$ charges.  To avoid this
anomaly it is required that in compact spaces these non-BPS branes
must exist in pairs~\cite{Uranga0011048}.  Considering a Type I
non-BPS D7-brane ($\widehat{\textrm{D7}}$-brane),  we know that it is
regarded as a pair of D7-brane and its world-sheet parity image
$\overline{\textrm{D7}}$-brane in Type IIB theory.  Thus we
require even numbers of these brane pairs in both Type IIA and IIB
theories, since Type IIA theory is the T-dual of Type IIB theory. 
 We only consider the effects from D3- and D7-branes since they do not
contribute the standard RR charges which have been considered
in the RR tadpole cancellation conditions.  The K-theory conditions for a
$\mathbf{\Z_2\times \Z_2}$ orientifold, which are 
equivalent to the global cancellations
of $\mathbf{\Z_2}$ RR charges carried by the D5$_i$-$\overline{\textrm{D5}}_i$
and D9$_i$-$\overline{\textrm{D9}}_i$ brane pairs,
were derived in~\cite{MS} and are given by
\begin{equation}
\sum_a Q9_a = \sum_a Q5_{1a} =  \sum_a Q5_{2a} = \sum_a Q5_{3a} =
0 \textrm{ mod }4 \label{K-charges}~.~\,
\end{equation}

\subsection{The Generalized Green-Schwarz Mechanism}

Although the cubic non-Abelian anomalies in intersecting D-brane 
models are cancelled automatically when the RR tadpole cancellation
 conditions are satisfied, the additional mixed $U(1)$ anomalies 
may still be present.  For
instance, the mixed $U(1)$-gravitational anomalies which generate masses to
the $U(1)$ gauge fields are not trivially zero~\cite{CSU2,CLS1,CLS2}. 
These anomalies are
cancelled by a generalized G-S mechanism which
involves the untwisted RR forms.  The couplings of the four
untwisted RR forms $B^i_2$ to the $U(1)$ field strength
$F_a$ of each stack $a$ are~\cite{IBII}
\begin{eqnarray}
 N_a m^1_a n^2_a n^3_a \int_{M4}B^1_2\wedge \textrm{tr}F_a~,~  \;\;
 N_a n^1_a m^2_a n^3_a \int_{M4}B^2_2\wedge \textrm{tr}F_a~,~
  \nonumber \\
 N_a n^1_a n^2_a m^3_a \int_{M4}B^3_2\wedge \textrm{tr}F_a~,~  \;\;
-N_a m^1_a m^2_a m^3_a \int_{M4}B^4_2\wedge \textrm{tr}F_a~.~\,
\end{eqnarray}
These couplings determine the linear combinations of $U(1)$ gauge
bosons that acquire string scale masses via the G-S mechanism.  
Sometimes, one combination of $U(1)$  gauge symmetries,
for example $U(1)_{I_{3R}}$, must remain as an exact 
gauge symmetry because
it is needed to generate the SM hypercharge interaction.
Therefore, we must ensure that the $U(1)_{I_{3R}}$ gauge boson 
 does not obtain such a mass. Suppose that the $U(1)_{I_{3R}}$ 
gauge symmetry is a linear
combination of the $U(1)$s :
\begin{equation}
U(1)_{I_{3R}}=\sum_a c_a U(1)_a~.~\,
\end{equation}
Because the corresponding field strength must be orthogonal to those that
acquire masses via the G-S mechanism, we have
\begin{eqnarray}
\sum_a c_a Q5_{1a} =0~,~& \;\; &\sum_a c_a Q5_{2a} =0  ~,~ \nonumber \\
\sum_a c_a Q5_{3a} =0~,~& \;\; &\sum_a c_a Q9_a =0 ~.~
\label{GSeq}
\end{eqnarray}

\section{Type IIA Model Building}

In this Section, we shall construct the trinification model and
the Pati-Salam like models on Type IIA orientifolds without flux.

\subsection{Trinification Model}

The $SU(3)_C\times SU(3)_L\times SU(3)_R$ trinification model, as
a candidate for a grand unified theory, was proposed by de
R\'{u}jula, Georgi, and Glashow~\cite{RGG} (see also~\cite{Babu:1985gi}).
 Although no one has considered such models,
the trinification model is quite interesting for 
the intersecting D-brane model building because all the left-handed
quarks $Q^i_L$, the right-handed quarks $Q^i_R$, the leptons $L^i$,
and the Higgs fields $H^k$, which are listed in 
Table \ref{trinificationI}, belong to the bi-fundamental representations.

\begin{table}[h]
\renewcommand{\arraystretch}{1.5}
\center
\begin{tabular}{|c||c|}
\hline

Particles & Representation   \\ \hline \hline

$Q^i_L$ &  $(\mathbf{3}, \mathbf{\bar{3}}, \mathbf{1})$ \\ \hline

$Q^i_R$ &  $(\mathbf{\bar{3}}, \mathbf{1}, \mathbf{3})$  \\ \hline

$L^i$~or~$H^k$ & $(\mathbf{1}, \mathbf{3}, \mathbf{\bar{3}})$ \\
\hline

\end{tabular}
\caption{The particle contents in the $SU(3)_C\times SU(3)_L\times SU(3)_R$ model.}
\label{trinificationI}
\end{table}

Let us briefly review the trinification model.
The electric charge generator $Q_{EM}$ is given by
\begin{equation}
Q_{EM} \equiv I_{3L} + \frac{Y}{2}
=I_{3L} - \frac{Y_L}{2} + I_{3R} - \frac{Y_R}{2}~,~\,
\end{equation}
where the generators for $U(1)_{I_{3L}}$ and $U(1)_{I_{3R}}$,
and $U(1)_{Y_L}$ and $U(1)_{Y_R}$  in 
$SU(3)_L$ and $SU(3)_R$ gauge symmetries are
\begin{equation}
\mathbf{T}_{U(1)_{I_{3L,R}}}= \left(  \begin{array}{ccc}
\frac{1}{2} & 0 & 0 \\
0 & -\frac{1}{2} & 0 \\
0 & 0 & 0  \end{array} \right)~,~\,
\end{equation}
\begin{equation}
\mathbf{T}_{U(1)_{Y_{L,R}}}= \left(  \begin{array}{ccc}
\frac{1}{3} & 0 & 0 \\
0 & \frac{1}{3} & 0 \\
0 & 0 & -\frac{2}{3}  \end{array} \right)~.~\,
\end{equation}

And the explicit particle components in the 
$(\mathbf{3}, \mathbf{\bar{3}}, \mathbf{1})$,
$(\mathbf{\bar{3}}, \mathbf{1}, \mathbf{3})$,
and $(\mathbf{1}, \mathbf{3}, \mathbf{\bar{3}})$
representations are
\begin{equation}
(\mathbf{3}, \mathbf{\bar{3}}, \mathbf{1}) :             \;\;
Q^i_L= \left(\begin{array}{ccc}
d & u & h \\
d & u & h \\
d & u & h  \end{array} \right) ~,~\,
\end{equation}

\begin{equation}
(\mathbf{\bar{3}}, \mathbf{1}, \mathbf{3}) :             \;\;
Q^i_R= \left(\begin{array}{ccc}
d^c & d^c & d^c \\
u^c & u^c & u^c \\
h^c & h^c & h^c  \end{array} \right) ~,~\,
\end{equation}

\begin{equation}
(\mathbf{1}, \mathbf{3}, \mathbf{\bar{3}}) :             \;\; L^i
\; \mathrm{or} \; H^k = \left(\begin{array}{ccc}
N & E^c & \nu \\
E & N^c & e \\
\nu^c & e^c & S  \end{array} \right) ~.~\,
\end{equation}

The $SU(3)_C\times SU(3)_L\times SU(3)_R$ gauge symmetry can be
broken down to the SM gauge symmetry by giving the vacuum
expectation values (VEVs) to $\nu^c$ and  $S$, {\it i.~e.},
\begin{equation}
\langle \nu^c \rangle \neq 0~,~ \;\; \langle S \rangle \neq 0 ~.~\,
\end{equation}

The electric charges for $h$ and $h^c$ are
respectively $-\frac{1}{3}$ and
$\frac{1}{3}$, for $E$ and $E^c$ are respectively $-1$
 and 1; and for $N$, $N^c$, and
$S$ are zero.

With above background, we can construct an intersecting D6-brane 
trinification model.
The bi-fundamental representation
with one fundamental and one anti-fundamental indices is
different from the  bi-fundamental representation with
 two  fundamental (or anti-fundamental) indices, for example, 
$(\mathbf{3, \overline{3}, 1})$ and $(\mathbf{3, 3, 1})$ 
(or $(\mathbf{\overline{3}, \overline{3}, 1})$).
So, we can contruct the trinification models with 
three families of the SM fermions and without tilted
two-torus.

There are three $SU(3)$ groups in the trinification model, so 
three stacks of six D6-branes are required. Additional stacks with
$U(1)$ group and filler branes are also used to satisfy the
RR tadpole cancellation conditions. 
In our model building,  we require
the intersection numbers to satisfy
\begin{equation}
I_{ab}=3 ; \;\; I_{ac}=-3 ; \;\; I_{bc} \geq 4~,~\,
\end{equation}
where $I_{ab}=3$ and  $I_{ac}=-3$  give us
three families of the left-handed quarks and
three families of the right-handed quarks, respectively,
and $I_{bc} \geq 4$ gives us  
three families of the leptons and ($I_{bc}-3$) Higgs field(s).

 We have large RR charges from three $SU(3)$ groups,
so it is not easy to construct a trinification
model without RR tadpoles. 
After careful searches, we find a supersymmetric intersecting D6-brane 
trinification model which satisfies the RR tadpole cancellation conditions
and K-theory conditions.
We present its complete wrapping numbers  and intersection numbers
in Table~\ref{tri-wrapping} 
and its spectrum  in Table~\ref{tri-spectrum} in Appendix A.
In this model, we have
three families of the SM fermions including the
right-handed neutrinos, one pair of Higgs doublets,
$H_u$ and $H_d$, 
one field $\nu^c$ and one field $S$.

\begin{table}[h]
\small
\begin{tabular}{|@{}c@{}|c||c@{}c@{}c||c|c||c|c|c|@{}c@{}|@{}c@{}|@{}c@{}|@{}c@{}|@{}c@{}|@{}c@{}|@{}c@{}|}
\hline

stack& $N_a$ & ($n_1$, $l_1$) & ($n_2$, $l_2$) & ($n_3$, $l_3$) &
A & S & $b$ & $b'$ & $c$ & $c'$ & $d$ & $d'$ & $e$ & $e'$ &
$f^{(1)}$ & $f^{(3)}$  \\ \hline \hline

 $a$ & 6 & ( 0, 1) & (-1,-1) & ( 2, 1) & -2 & 2 & 3 & 1 & -3 &
-1 & -3 & -3 & -3 & -3 & -2 & 0  \\ \hline

 $b$ & 6 & (-1,-1) & (-2, 1) & ( 1, 0) &  2 & -2 & - & - & 4 &
0(2) & 6 & 0(5) & 6 & 0(5) & 2 & -1   \\ \hline

 $c$ & 6 & (-1, 1) & ( 0, 1) & (-1, 1) & 0 & 0 & - & - & - &
- & 0(-2) & 0(2) & 0(-2) & 0(2) & 0 & 1   \\ \hline

 $d$ & 2 & (-1, 1) & (-1, 2) & ( 1, 1) & -16 & 0 & - & - & - &
- & - & - & 0(0) & -16 & -1 & -2   \\ \hline

 $e$ & 2 & (-1, 1) & (-1, 2) & ( 1, 1) & -16 & 0 & - & - & - &
- & - & - & - & - & -1 & -2   \\ \hline \hline

$fil^{(2)}$ & 2 & ( 1, 0) & ( 0, 1) & ( 0,-1) & - & - & - & - & -
& - & - & - & - & - & - & -  \\ \hline

$fil^{(3)}$ & 6 & ( 0, 1) & ( 1, 0) & ( 0,-1) & - & - & - & - & -
& - & - & - & - & - & - & -
\\ \hline
\end{tabular}
\caption{Wrapping numbers and intersection numbers in the 
$SU(3)_C\times SU(3)_L\times SU(3)_R$ model.} \label{tri-wrapping}
\end{table}

(1) Gauge Symmetry Breaking

The $U(3)_C\times U(3)_L\times U(3)_R$ gauge symmetry is
broken down to the $SU(3)_C\times SU(3)_L\times SU(3)_R$ gauge symmetry
due to the G-S mechanism. And
the $SU(3)_C\times SU(3)_L\times SU(3)_R$ gauge symmetry can be
broken down to the $SU(3)_C\times SU(2)_L\times U(1)_{Y_L}\times
U(1)_{I_{3R}}\times U(1)_{Y_R}$ gauge symmetry by the splittings
of the $U(3)_L$ and $U(3)_R$ stacks of the D6-branes.
 Giving VEVs to the singlet Higgs fields $\nu^c$ and  $S$, we can break the 
 $U(1)_{Y_L}\times U(1)_{I_{3R}}\times U(1)_{Y_R}$ gauge symmetry
down to the $U(1)_Y$ hypercharge interaction.  The complete gauge symmetry
breaking chains are 
\begin{eqnarray}
& & SU(3)_C \times SU(3)_L\times SU(3)_R       \nonumber         \\
& \stackrel{Splitting}{\longrightarrow} & SU(3)_C\times
SU(2)_L\times U(1)_{Y_L}\times
U(1)_{I_{3R}}\times U(1)_{Y_R}               \nonumber       \\
& \stackrel{VEVs}{\longrightarrow} & SU(3)_C\times SU(2)_L\times
U(1)_Y ~.~\,
\end{eqnarray}

We assume the low scale supersymmetry in this paper. Then
the VEVs for $\nu^c$ and $S$ should be around the TeV scale because their Higgs
mechanism can not preserve the D-flatness and F-flatness and then
breaks four-dimensional $N=1$ supersymmetry.

(2) Fermion Masses and Mixings

The quark Yukawa couplings $ y_{ijk} Q^i_L Q^j_R H^k$ 
are allowed by the anomalous $U(1)$ gauge symmetries in
the intersecting D6-brane trinification model, while
the lepton and neutrino Yukawa couplings $ y'_{ijk} L_i L_j H^k$ 
 are forbidden by the anomalous $U(1)_L\times
U(1)_R \subset U(3)_L\times U(3)_R$ gauge symmetry.

In our model, only one family of the SM quarks can obtain masses because
  $Q^i_L$ arise from the
intersections on the second two-torus, while $Q^i_R$ arise from the
intersections on the third two-torus, or because we only have 
one pair of Higgs doublet fields.

\subsection{Pati-Salam Model}

In the previous model building with or without fluxes, it is very difficult
 to generate suitable three-family SM fermion masses
and mixings.  In the $SU(5)$ models  and flipped $SU(5)$ models, the up-type
quark Yukawa couplings and the down-type quark Yukawa couplings are 
forbidden by 
anomalous $U(1)$ gauge symmetries. And for the Pati-Salam like models,
although all the Yukawa couplings could be allowed in principle, 
it is very difficult to construct three-family models which can give 
 suitable masses and mixings to three families of the SM fermions
because the left-handed fermions,
the right-handed fermions and the Higgs fields in general arise from 
the intersections on different two-tori~\cite{CLLL-P}. 
 Moreover, if supersymmetry is broken by supergravity fluxes, 
it seems that the masses for the massless SM fermions 
may not be generated from radiative corrections~\cite{CLLL-P}
 because the supersymmetry breaking trilinear soft terms are 
universal and the supersymmetry breaking soft masses for the left/right-chiral
squarks and sleptons are also 
universal~\cite{Soft BreakingI,Soft BreakingII,Soft BreakingIII}.
Thus, how to construct the
Standard-like models, which can give suitable fermion
masses and mixings for three families, is an interesting problem.

To solve this problem, we construct another class of
supersymmetric  Pati-Salam models without RR tadpoles and
K-theory anomaly. In particular,
under the $U(4)_C \times U(2)_L \times U(2)_R$ gauge symmetry,
 we consider
the particles with quantum numbers $(\mathbf{4, \overline{2}, 1})$
and $(\mathbf{ \overline{4}, 1, 2})$  as
the SM fermions while we consider the
particles with quantum numbers $(\mathbf{ 4, 2, 1})$
and $(\mathbf{ \overline{4}, 1, \overline{2}})$ as exotic particles
because these particles are distinguished by anomalous $U(1)$ gauge 
symmetries (In the trinification models, these kinds of particles
are obviously different.). With this convention,
we can construct the three-family Pati-Salam models without 
tilted two-torus where the left-handed and right-handed
SM fermions and the Higgs fields arise from the intersections on the
same two-torus, and then, we may explain three-family
SM fermion masses and mixings.

In this kind of Pati-Salam model building, we require the
intersection numbers to satisfy
\begin{equation}
I_{ab} = 3; \;\; I_{ac} = -3; \;\; I_{bc}\geq 1~,~
\label{PS-ISN}
\end{equation}
where $I_{ab}=3$ and  $I_{ac}=-3$  give us
three families of the left-handed fermions and
three families of the right-handed fermions, respectively,
and $I_{bc} \geq 1$ gives us  
bidoublet Higgs field(s) with allowed Yukawa couplings.

We present one concrete model whose  wrapping numbers 
and intersection numbers are given
in Table \ref{422-wrapping2}. In this model,
 the absolute values of the intersection numbers on the seond two-torus  
between the $U(4)_C$ and  $U(2)_L$ stacks of D6-branes,
 between the $U(4)_C$ and  $U(2)_R$ stacks of D6-branes, 
and between the $U(2)_L$ and $U(2)_R$ stacks of D6-branes
are all three, and all the Yukawa couplings are allowed by anomalous
$U(1)$ gauge symmetries.  Therefore, we may explain the  masses and mixings
for three families of the SM fermions.  Note that we have four
additional D6-brane stacks $d$, $e$, $f$ and $g$
  in this model, for which one can arbitrarily
substitute them into filler brane stacks ($USp$ groups).

In general, the $U(4)_C \times U(2)_L \times U(2)_R$ gauge symmetry
can be broken down to the 
$SU(3)_C \times SU(2)_L \times U(1)_{I_{3R}} \times U(1)_{B-L}$ 
gauge symmetry
due to the G-S mechanism and the splittings of the
$U(4)_C$ and $U(2)_R$ stacks of D6-branes. 
In our  model, the $U(1)_{I_{3R}} \times U(1)_{B-L}$
gauge symmetry can only be broken down to the $U(1)_Y$ gauge
symmetry by giving VEVs to the scalar components of the
right-handed neutrino superfields or the neutral component
in the multiplet $(\mathbf{ \overline{4}, 1, \overline{2}})$
from $I_{ac'}$ intersection.
However, this Higgs mechanism  can not
preserve the D-flatness and F-flatness, and then
breaks four-dimensional $N=1$ supersymmetry.
Therefore, the $U(1)_{I_{3R}} \times U(1)_{B-L}$
gauge symmetry breaking scale should be around 
the TeV scale.

\begin{table}[h]
\small
\begin{tabular}{|@{}c@{}|@{}c@{}||@{}c@{}c@{}c@{}||c|c||c|c|c|@{}c@{}|
@{}c@{}|@{}c@{}|@{}c@{}|@{}c@{}|@{}c@{}|@{}c@{}|@{}c@{}|@{}c@{}|}
\hline

stack& $N_a$ & ($n_1$, $l_1$) & ($n_2$, $l_2$) & ($n_3$, $l_3$) &
A & S & $b$ & $b'$ & $c$ & $c'$ & $d$ & $d'$ & $e$ & $e'$ & $f$ &
$f$ & $g$ & $g'$  \\ \hline \hline

 $a$ & 8 & (-1, 0) & (-1, 1) & ( 1, 1) & 0 & 0 & 3 & 1 & -3 &
-1 & -3 & -1 & -3 & -1 & 1 & 3 & 1 & 3 \\ \hline

 $b$ & 4 & ( 1,-1) & ( 1, 2) & ( 1, 0) &  2 & -2 & - & - & 6 &
0(5) & 8 & 0(4) & 8 & 0(4) & -2 & 0(1) & -2 & 0(1)  \\ \hline

 $c$ & 4 & ( 1, 1) & ( 2, 1) & ( 0,-1) & -2 & 2 & - & - & - &
- & 0(1) & 2 & 0(1) & 2 & 0(-4) & -8 & 0(-4) & -8   \\ \hline

 $d$ & 2 & (-1,-1) & (-1, 0) & ( 1,-2) & -2 & 2 & - & - & - &
- & - & - & 0(0) & 0(0) & 0(5) & -6 & 0(5) & -6   \\ \hline

 $e$ & 2 & (-1,-1) & (-1, 0) & ( 1,-2) & -2 & 2 & - & - & - &
- & - & - & - & - & 0(5) & -6 & 0(5) & -6   \\ \hline

 $f$ & 2 & ( 1, 1) & ( 0, 1) & (-2,-1) & -2 & 2 & - & - & - & - & -
& - & - & - & - & - & 0(0) & 0(0) \\ \hline

 $g$ & 2 & ( 1, 1) & ( 0, 1) & (-2,-1) & -2 & 2 & - & - & - & - & -
& - & - & - & - & - & - & -  \\ \hline

\end{tabular}
\caption{Wrapping numbers and intersection numbers in the 
$U(4)_C\times U(2)_L\times U(2)_R\times U(1)^4$ Model. } 
\label{422-wrapping2}
\end{table}

\subsection{$U(4)_C\times U(2)_L\times U(1)' \times U(1)''$ Model}

In all the previous Pati-Salam like model building, the
$U(1)_{I_{3R}}$ arises from the non-Abelian gauge symmetry, for
example, $U(2)_R$ or $USp(2f)_R$.
However, $U(1)_{I_{3R}}$ may come from a linear combination of 
 $U(1)$ gauge symmetries.

 In our model building, we require
\begin{equation}
I_{ab} = 3~;~ \;\; I_{ac}=I_{ad}=-3~;~ \;\; I_{bc}\geq 1~;~ \;\;
I_{bd}\geq1 ~,~\,
\end{equation}
where $I_{ab}=3$ and $I_{ac}=I_{ad}=-3$  give us
three families of the left-handed fermions and
three families of the right-handed fermions, respectively,
and $I_{bc} \geq 1$ gives us  
bidoublet Higgs fields with allowed Yukawa couplings.

Let us give a concrete supersymmetric model without the RR tadpoles and
K-theory anomaly. We present
the wrapping numbers and intersecting numbers 
in Table \ref{4211-wrapping}, and the spectrum 
in Tables \ref{4211-spectrum-A} and \ref{4211-spectrum-B} in Appendix A.  
In particular, the c and d stacks of D6-branes are not T-dual
to each other (If c and d stacks of D6-branes are  T-dual
to each other, the gauge symmetry in fact is $U(4)_C\times U(2)_L\times U(2)_R$.). 
There are totally six $U(1)$ gauge symmetries
where four combinations of them are global and their gauge fields
obtain masses by the G-S mechanism. The rest two
combinations, which are the massless anomaly-free $U(1)_{I_{3R}}$ and
 $U(1)_{X}$ gauge symmetries, are given by
\begin{eqnarray}
U(1)_{I_{3R}}&=& \frac{1}{2}(U(1)_a+U(1)_b+2U(1)_c) ~,~\\
U(1)_X &=& \frac{1}{2} (U(1)_a-U(1)_b+2U(1)_d-2U(1)_e-2U(1)_f)~.~\,
\end{eqnarray}
In addition, 
the $U(4)_C \times U(2)_L \times U(1)' \times U(1)'' \times U(1)_e \times U(1)_f$
gauge symmetry can be broken down to the
$SU(3)_C \times SU(2)_L \times U(1)_{I_{3R}} \times U(1)_{B-L} \times U(1)_{X}$
gauge symmetry by the G-S mechanism and 
the splitting of the $U(4)_C$ stack of D6-branes. 
Furthermore, the $U(1)_{X}$ gauge symmetry can be broken by giving VEVs
to the SM singlets $1_d$ and $1_e$ (or $1_f$) which
are charged under $U(1)_{X}$ 
(see the spectrum in Table \ref{4211-spectrum-B} in Appendix A). And
the $U(1)_{I_{3R}} \times U(1)_{B-L}$
gauge symmetry can be broken down to the $U(1)_Y$ gauge
symmetry by giving VEVs to $(\bar{4}_a,1_e)$ (or $(\bar{4}_a,\bar{1}_e)$)
and $(4_a,1_f)$ (or $(4_a,\bar{1}_f)$).
Because these Higgs mechanism can keep the D-flatness and
F-flatness and then preserve four-dimensional $N=1$ 
 supersymmetry, these
gauge symmetry breaking scales can be close to the
string scale. However, only
 one family of the SM fermions can obtain masses.

\begin{table}[h]
\small
\begin{tabular}{|@{}c@{}|@{}c@{}||@{}c@{}c@{}c@{}||c|c||c|c|c|@{}c@{}|
c|c|@{}c@{}|@{}c@{}|@{}c@{}|c|@{}c@{}|@{}c@{}|} \hline

stack& $N_a$ & ($n_1$, $l_1$) & ($n_2$, $l_2$) & ($n_3$, $l_3$) &
A & S & $b$ & $b'$ & $c$ & $c'$ & $d$ & $d'$ & $e$ & $e'$ & $f$ &
$f'$ & $fil^{(1)}$ & $fil^{(3)}$  \\ \hline \hline

 $a$ & 8 & (-1, 0) & (-1, 1) & ( 1, 1) & 0 & 0 & 3 & 1 & -3 &
-1 & -3 & -1 & -3 & -1 & 1 & 3 & -1 & 1  \\ \hline

 $b$ & 4 & ( 1,-1) & ( 1, 2) & ( 1, 0) &  2 & -2 & - & - & 8 &
0(4) & 9 & -5 & 8 & 0(4) & -3 & -1 & 2 & 0  \\ \hline

 $c$ & 2 & ( 1, 1) & ( 1, 0) & ( 1,-2) & -2 & 2 & - & - & - &
- & 1 & 3 & 0(0) & 0(0) & 5 & -9 & 0 & -2   \\ \hline

 $d$ & 2 & ( 2, 1) & ( 2, 1) & ( 0,-1) & -6 & 6 & - & - & - &
- & - & - & -1 & 3 & 0(-4) & -16 & 0 & -4   \\ \hline

 $e$ & 2 & (-1,-1) & (-1, 0) & ( 1,-2) & -2 & 2 & - & - & - &
- & - & - & - & - & 5 & -9 & 0 & -2   \\ \hline

 $f$ & 2 & ( 2, 1) & ( 0,-1) & ( 2, 1) & -6 & 6 & - & - & - &
- & - & - & - & - & - & - & -4 & 0   \\ \hline \hline

$fil^{(3)}$ & 4 & ( 0, 1) & ( 1, 0) & ( 0,-1) & - & - & - & - & -
& - & - & - & - & - & - & - & - & - \\ \hline

$fil^{(4)}$ & 4 & ( 0, 1) & ( 0,-1) & ( 1, 0) & - & - & - & - & -
& - & - & - & - & - & - & - & - & - \\ \hline

\end{tabular}
\caption{Wrapping numbers and intersection numbers in the 
$U(4)_C\times U(2)_L\times U(1)' \times U(1)'' \times U(1)_e \times U(1)_f
\times USp(4)\times
USp(4)$ model.} \label{4211-wrapping}
\end{table}

\section{Type IIB Flux Model Building}

In this Section, we shall consider the trinification models, and
the Pati-Salam like models on Type IIB orientifold with flux
compactifications, which are very interesting because
 the supergravity fluxes can 
stabilize the dilaton and the complex structure parameters.

 For the trinification models, we already have quite large RR charges due to 
the three $SU(3)$ groups. With  Type IIB supergravity fluxes,
the RR tadpole cancellation conditions are much more difficult to be
satisfied. And in our detail calculations, we find that it may be
  impossible to find such a model. 

For the Pati-Salam like models,  the  three-family and four-family 
Standard-like models with one unit of quantized flux and with
the electroweak sector from $USp$ groups were obtained~\cite{MS,CL}
by introducing  magnetized D9-branes with large negative D3-brane
charges in the hidden sector, and many supersymmtric and 
non-supersymmetric $U(4)_C \times U(2)_L \times U(2)_R$
models were constructed by considering 
the magnetized D9-branes with large negative D3-brane charges 
in the SM observable sector~\cite{Cvetic:2005bn}.
Here, we consider a new flux model with 
$U(4)_C \times U(2)_L \times U(2)_R$ gauge symmetry where the
 magnetized D9-branes with large negative D3-brane
charges are introduced in the hidden sector. This kind of models
has not been studied previously because it is very difficult to
have supersymmetric D-brane configurations with more than three stacks
of $U(n)$ branes.

In the model building, we require the intersection numbers to 
satisfy the conditions in Eq. (\ref{PS-ISN}).
We find a model with one unit of flux,
and its wrapping numbers and intersection numbers are given in Table
\ref{422-wrappingflux}.  Interestingly, no filler branes are
needed so we do not have any $USp$ groups.  The two extra $U(1)$
gauge symmetries are utilized to compensate 
the large positive D3-brane charges due to the
supergravity fluxes.  
The $U(4)_C \times U(2)_L \times U(2)_R$ gauge symmetry
can be broken down to the 
$SU(3)_C \times SU(2)_L \times U(1)_{I_{3R}} \times U(1)_{B-L}$ 
gauge symmetry
by the G-S mechanism and the splittings of the
$U(4)_C$ and $U(2)_R$ stacks of D6-branes.

However, the $U(1)_{I_{3R}} \times U(1)_{B-L}$
gauge symmetry can only be broken down to the $U(1)_Y$ gauge
symmetry at the TeV scale by giving VEVs to the scalar components of the
right-handed neutrino superfields or the neutral component
in the multiplet $(\mathbf{ \overline{4}, 1, \overline{2}})$
from $I_{ac'}$ intersection because this Higgs mechanism  can not
preserve the D-flatness and F-flatness, and then
breaks four-dimensional $N=1$ supersymmetry.
Also, we can only give 
masses to one family of the SM fermions.

\begin{table}[h]
\small
\begin{tabular}{|@{}c@{}|@{}c@{}||@{}c@{}c@{}c@{}||c|c||c|c|c|@{}c@{}|c|c|@{}c@{}|@{}c@{}|} \hline

stack& $N_a$ & ($n_1$, $l_1$) & ($n_2$, $l_2$) & ($n_3$, $l_3$) &
A & S & $b$ & $b'$ & $c$ & $c'$ & $d$ & $d'$ & $e$ & $e'$  \\
\hline \hline

 $a$ & 8 & (-1, 0) & (-1, 1) & ( 1, 1) & 0 & 0 & 3 & 1 & -3 &
-1 & 3 & -3 & 3 & -3  \\ \hline

 $b$ & 4 & ( 1,-1) & ( 1, 2) & ( 1, 0) &  2 & -2 & - & - & 8 &
0(4) & 9 & -5 & 9 & -5  \\ \hline

 $c$ & 4 & ( 1, 1) & ( 1, 0) & ( 1,-2) & -2 & 2 & - & - & - &
- & 5 & -9 & 5 & -9   \\ \hline

 $d$ & 2 & ( 2, 1) & (-2,-1) & ( 2, 1) & -54 & -10 & - & - & - &
- & - & - & 0(0) & -64   \\ \hline

 $e$ & 2 & ( 2, 1) & (-2,-1) & ( 2, 1) & -54 & -10 & - & - & - &
- & - & - & - & -  \\ \hline \hline

\end{tabular}
\caption{Wrapping numbers and intersection numbers in the 
$U(4)_C\times U(2)_L\times U(2)_R \times U(1)^2$ model with one unit of
flux.} \label{422-wrappingflux}
\end{table}

\section{Discussions and Conclusions}

We have constructed new Standard-like models on Type II orientifolds.
In Type IIA theory on $\mathbf{T^6/(\Z_2\times \Z_2)}$
 orientifold with intersecting D6-branes,
we first constructed a three-family trinification model
where all the SM fermions and Higgs fields belong to the bi-fundamental
representations.
The $U(3)_C\times U(3)_L\times U(3)_R$ gauge symmetry
can be broken down to  the $SU(3)_C\times SU(2)_L\times U(1)_{Y_L}\times
U(1)_{I_{3R}}\times U(1)_{Y_R}$ gauge symmetry by
the G-S mechanism and D6-brane splittings, and further down
to the SM gauge symmetry  at the TeV
scale by the Higgs mechanism which
 can not preserve the D-flatness and F-flatness and then
breaks four-dimensional $N=1$ supersymmetry. In the general intersecting
D-brane trinification models, 
the quark Yukawa couplings are allowed while the
lepton Yukawa couplings are forbidden by anomalous $U(1)$ gauge symmetries.
And in our model, we can only give masses to one
family of the SM quarks.
In addition, we  constructed a Pati-Salam model
which may generate suitable three-family SM fermion masses and 
mixings. The $U(4)_C \times U(2)_L \times U(2)_R$ gauge symmetry
can be broken down to the 
$SU(3)_C \times SU(2)_L \times U(1)_{I_{3R}} \times U(1)_{B-L}$ 
gauge symmetry due to the G-S mechanism and D6-brane splittings. 
In our  model, the $U(1)_{I_{3R}} \times U(1)_{B-L}$
gauge symmetry can only be broken down to the $U(1)_Y$ gauge
symmetry at the TeV scale by Higgs mechanism.
Moreover, we constructed for the first time  a Pati-Salam like model with  
 $U(4)_C \times U(2)_L \times U(1)' \times U(1)''$ gaue symmetry 
where the $U(1)_{I_{3R}}$  gauge symmetry comes from a linear combination of
  $U(1)$ gauge symmetries. And  
 the $U(1)_{I_{3R}} \times U(1)_{B-L}$
gauge symmetry can  be broken down to the $U(1)_Y$ gauge
symmetry at (or close to) the string scale by Higgs mechanism.
However, only one family of the SM fermions can obtain masses.

In Type IIB theory on $\mathbf{T^6/(\Z_2\times \Z_2)}$ 
 orientifold with flux compactifications, 
 we could not construct the trinification
model with supergravity fluxes because the three $SU(3)$ groups
already contribute large RR charges. 
We constructed a new flux model with 
$U(4)_C \times U(2)_L \times U(2)_R$ gauge symmetry where the
 magnetized D9-branes with large negative D3-brane
charges are introduced in the hidden sector. This kind of models
has not been studied previously because it is very difficult to
have supersymmetric D-brane configurations with more than three stacks
of $U(n)$ branes. However, in our model,
 the $U(1)_{I_{3R}} \times U(1)_{B-L}$
gauge symmetry can only be broken down to the $U(1)_Y$ gauge
symmetry at the TeV scale, and we can only give 
masses to one family of the SM fermions.

\section*{Acknowledgments}
We would like to thank M.~Cveti\v c for helpful discussions.
T.L. is grateful to the George P. and Cynthia W. Mitchell 
Institute for Fundamental Physics for hospitality during 
the early stage of this project.
The research of T.L. was supported by DOE grant DE-FG02-96ER40959, 
and the research of D.V.N. was supported by DOE grant 
DE-FG03-95-Er-40917.

\newpage

\appendix

\section{Spectra}

In this Appendix, we present the spectrum in the $SU(3)_C\times SU(3)_L\times
SU(3)_R\times U(1)^5\times USp(2)\times USp(6)$ model with four
global $U(1)$s from the G-S mechanism in Table \ref{tri-spectrum}, and the 
spectrum in the $U(4)_C\times U(2)_L\times U(1)' \times
U(1)'' \times U(1)_e \times U(1)_f
\times USp(4)\times USp(4)$ model with anomaly free
 $U(1)_{I_{3R}}$ and $U(1)_{X}$ gauge symmetries
in Tables \ref{4211-spectrum-A} and \ref{4211-spectrum-B}.

\begin{table}[h]

\begin{center}
\small
\begin{tabular}{|c||@{}c@{}||@{}c@{}|@{}c@{}|@{}c@{}|@{}c@{}|@{}c@{}||@{}c@{}|
@{}c@{}|@{}c@{}|@{}c@{}|} \hline

 Rep. & Multi. &$U(1)_a$&$U(1)_b$&$U(1)_c$& $U(1)_d$
& $U(1)_e$ & $U(1)_1$ & $U(1)_2$ &
$U(1)_3$ & $U(1)_4$   \\
\hline \hline

$(3_a,\bar{3}_b)$ & 3 & 1 & -1 & 0 & 0 & 0 & -24 & 6 & 0 & 6  \\

$(\bar{3}_a,3_c)$ & 3 & -1 & 0 & 1 & 0 & 0 & 12 & 6 & 0 & -12  \\

$(3_b,\bar{3}_c)$ & 3 & 0 & 1 & -1 & 0 & 0 & 12 & -12 & 0 & 6  \\
 \hline

$(3_b,\bar{3}_c)$ & 1 & 0 & 1 & -1 & 0 & 0 & 12 & -12 & 0 & 6  \\
\hline

$(3_a,3_b)$       & 1 & 1 & 1 & 0 & 0 & 0 & 0 & -6 & 0 & 6  \\

$(\bar{3}_a,\bar{3}_c)$ & 1 & -1 & 0 & -1 & 0 & 0 & 12 & -6 & 0 & 0  \\

$(\bar{3}_a,1_d)$ & 3 & -1 & 0 & 0 & 1 & 0 & 10 & -4 & 2 & -10  \\

$(\bar{3}_a,\bar{1}_d)$ & 3 & -1 & 0 & 0 & -1 & 0 & 14 & 4 & -2 & -2  \\

$(\bar{3}_a,1_e)$ & 3 & -1 & 0 & 0 & 0 & 1 & 10 & -4 & 2 & -10  \\

$(\bar{3}_a,\bar{1}_e)$ & 3 & -1 & 0 & 0 & 0 & -1 & 14 & 4 & -2 & -2  \\

$(3_b,\bar{1}_d)$ & 6 & 0 & 1 & 0 & -1 & 0 & 14 & -2 & -2 & 4  \\

$(3_b,\bar{1}_e)$ & 6 & 0 & 1 & 0 & 0 & -1 & -2 & -2 & 4 & 4  \\

$(\bar{1}_d,\bar{1}_e)$ & 16 & 0 & 0 & 0 & -1 & -1 & 4 & 8 & -4 & 8  \\

$A_a$ & 2 & -2 & 0 & 0 & 0 & 0 & 24 & 0 & 0 & -12  \\

$A_b$ & 2 & 0 & 2 & 0 & 0 & 0 & 24 & -12 & 0 & 0  \\

$S_a$ & 2 & 2 & 0 & 0 & 0 & 0 & -24 & 0 & 0 & 12  \\

$S_b$ & 2 & 0 & -2 & 0 & 0 & 0 & -24 & 12 & 0 & 0  \\

\hline

\multicolumn{11}{|c|}{Additional non-chiral and $USp(2)$ \& $USp(6)$ Matter}\\ \hline
\end{tabular}
\caption{The spectrum in the $SU(3)_C\times SU(3)_L\times
SU(3)_R\times U(1)^5\times USp(2)\times USp(6)$ model with four
global $U(1)$s from the G-S mechanism. }
\label{tri-spectrum}
\end{center}
\end{table}

\begin{table}[h]

\begin{center}
\small
\begin{tabular}{|c||@{}c@{}||@{}c@{}|@{}c@{}|@{}c@{}|@{}c@{}|@{}c@{}|@{}c@{}||
@{}c@{}|@{}c@{}|} \hline

 Rep. & Multi. &$U(1)_a$&$U(1)_b$&$U(1)_c$& $U(1)_d$
& $U(1)_e$ & $U(1)_f$ &  $2U(1)_{I_{3R}}$ & $2U(1)_X$    \\
\hline \hline

$(4_a,\bar{2}_b)$ & 3 & 1 & -1 & 0 & 0 & 0 & 0 & 0 & 2   \\

$(\bar{4}_a,1_c)$ & 3 & -1 & 0 & 1 & 0 & 0 & 0 &  1 & -1   \\

$(\bar{4}_a,1_d)$ & 3 & -1 & 0 & 0 & 1 & 0 & 0 & -1 & 1   \\
\hline
 $(2_b,\bar{1}_c)$ & 8 & 0 & 1 & -1 & 0 & 0 & 0 & -1 & -1   \\

$(2_b,\bar{1}_d)$ & 9 & 0 & 1 & 0 & -1 & 0 & 0 & 1 & -3   \\
\hline 
\end{tabular}
\caption{The SM fermions and Higgs fields in the 
$U(4)_C\times U(2)_L\times U(1)' \times U(1)'' \times U(1)_e \times U(1)_f
\times USp(4)\times USp(4)$ model, with anomaly free 
$U(1)_{I_{3R}}$ and $U(1)_{X}$ gauge symmetries. }
\label{4211-spectrum-A}
\end{center}
\end{table}

\begin{table}[h]
\begin{center}
\small
\begin{tabular}{|c||@{}c@{}||@{}c@{}|@{}c@{}|@{}c@{}|@{}c@{}|@{}c@{}|@{}c@{}||
@{}c@{}|@{}c@{}|} \hline
 Rep. & Multi. &$U(1)_a$&$U(1)_b$&$U(1)_c$& $U(1)_d$
& $U(1)_e$ & $U(1)_f$ &  $2U(1)_{I_{3R}}$ & $2U(1)_X$    \\
\hline \hline
$(4_a,2_b)$       & 1 & 1 & 1 & 0 & 0 & 0 & 0 & 2 & 0  \\

$(\bar{4}_a,\bar{1}_c)$ & 1 & -1 & 0 & -1 & 0 & 0 & 0 & -3 & -1   \\

$(\bar{4}_a,\bar{1}_d)$ & 1 & -1 & 0 & 0 & -1 & 0 & 0 & -1 & -3  \\

$(\bar{4}_a,1_e)$ & 3 & -1 & 0 & 0 & 0 & 1 & 0 & -1 & -3   \\

$(\bar{4}_a,\bar{1}_e)$ & 1 & -1 & 0 & 0 & 0 & -1 & 0 & -1 & 1   \\

$(4_a,\bar{1}_f)$ & 1 & 1 & 0 & 0 & 0 & 0 & -1 & 1 & 3   \\

$(4_a,1_f)$ & 3 & 1 & 0 & 0 & 0 & 0 & 1 & 1 & -1   \\

$(\bar{2}_b,\bar{1}_d)$ & 5 & 0 & -1 & 0 & -1 & 0 & 0 & -1 & -1  \\

$(2_b,\bar{1}_e)$ & 8 & 0 & 1 & 0 & 0 & -1 & 0 & 1 & 1   \\

$(\bar{2}_b,1_f)$ & 3 & 0 & -1 & 0 & 0 & 0 & 1 & -1 & -1   \\

$(\bar{2}_b,\bar{1}_f)$ & 1 & 0 & -1 & 0 & 0 & 0 & -1 & -1 & 3   \\

$(1_c,\bar{1}_d)$ & 1 & 0 & 0 & 1 & -1 & 0 & 0 & 2 & -2   \\

$(1_c,1_d)$ & 3 & 0 & 0 & 1 & 1 & 0 & 0 & 2 & 2   \\

$(1_c,\bar{1}_f)$ & 5 & 0 & 0 & 1 & 0 & 0 & -1 & 2 & 2   \\

$(\bar{1}_c,\bar{1}_f)$ & 9 & 0 & 0 & -1 & 0 & 0 & -1 & -2 & 2   \\

$(\bar{1}_d,1_e)$ & 1 & 0 & 0 & 0 & -1 & 1 & 0 & 0 & -4   \\

$(1_d,1_e)$ & 3 & 0 & 0 & 0 & 1 & 1 & 0 & 0 & 0   \\

$(\bar{1}_d,\bar{1}_f)$ & 16 & 0 & 0 & 0 & -1 & 0 & -1 & 0 & 0   \\

$(1_e,\bar{1}_f)$ & 5 & 0 & 0 & 0 & 0 & 1 & -1 & 0 & 0   \\

$(\bar{1}_e,\bar{1}_f)$ & 9 & 0 & 0 & 0 & 0 & -1 & -1 & 0 & 4   \\

$1_b$ & 2 & 0 & 2 & 0 & 0 & 0 & 0 & 2 & -2  \\

$\bar{3}_b$ & 2 & 0 & -2 & 0 & 0 & 0 & 0 & -2 & 2  \\

$1_c$ & 2 & 0 & 0 & 2 & 0 & 0 & 0 & 4 & 0  \\

$1_d$ & 6 & 0 & 0 & 0 & 2 & 0 & 0 & 0 & 4  \\

$1_e$ & 2 & 0 & 0 & 0 & 0 & 2 & 0 & 0 & -4  \\

$1_f$ & 6 & 0 & 0 & 0 & 0 & 0 & 2 & 0 & -4  \\
\hline
\multicolumn{10}{|c|}{Additional non-chiral and $USp(4)$ \& $USp(4)$ Matter}\\ \hline
\end{tabular}
\caption{The extra particles in the 
$U(4)_C\times U(2)_L\times U(1)' \times U(1)'' \times U(1)_e \times U(1)_f
\times USp(4)\times USp(4)$ model, with anomaly free 
$U(1)_{I_{3R}}$ and $U(1)_{X}$ gauge symmetries.  }
\label{4211-spectrum-B}
\end{center}
\end{table}


\begin{thebibliography}{99}
\itemsep 0.5mm


\bibitem{JPEW}
J.~Polchinski and E.~Witten, Nucl.\ Phys.\ B {\bf 460}, 525
(1996).

\bibitem{ABPSS}
C.~Angelantonj, M.~Bianchi, G.~Pradisi, A.~Sagnotti and
Y.~S.~Stanev, Phys.\ Lett.\ B {\bf 385}, 96 (1996).

\bibitem{berkooz}
M.~Berkooz and R.G.~Leigh, Nucl.\ Phys.\ B {\bf 483}, 187 (1997).

\bibitem{bdl}
M.~Berkooz, M.~R.~Douglas and R.~G.~Leigh, Nucl. Phys. B {\bf 480}
(1996) 265.

\bibitem{bachas}
C.~Bachas, hep-th/9503030.


\bibitem{LU}
R.~Blumenhagen, L.~G\"orlich, B.~K\"ors and D.~L\"ust, JHEP {\bf 0010} (2000)
006.

\bibitem{IB}
G.~Aldazabal, S.~Franco, L.~E.~Ib\'a\~nez, R.~Rabad\'an and A.~M.~Uranga, JHEP {\bf
0102}, 047 (2001).


\bibitem{IBII}
G.~Aldazabal, S.~Franco, L.~E.~Ib\'a\~nez, R.~Rabad\'an and A.~M.~Uranga, J.\ Math.\
Phys.\  {\bf 42}, 3103 (2001).

\bibitem{LUII}
R.~Blumenhagen, B.~K\"ors and D.~L\"ust, JHEP {\bf 0102} (2001) 030.

\bibitem{Sagnottietal}
C.~Angelantonj, I.~Antoniadis, E.~Dudas and A.~Sagnotti,
Phys.\ Lett.\ B {\bf 489}, 223 (2000).

\bibitem{CIM}
D.~Cremades, L.E.~Ib\'a\~nez and F.~Marchesano,
JHEP {\bf 0307}, 038 (2003).

\bibitem{List}
L.~E.~Ib\'a\~nez, F.~Marchesano and R.~Rabad\'an, JHEP {\bf 0111}, 002 (2001);
R.~Blumenhagen, B.~K\"ors and D.~L\"ust, T.~Ott, Nucl. Phys. {\bf B616} (2001) 3;
D.~Cremades, L.~E.~Ib\'a\~nez and F.~Marchesano, Nucl.\ Phys.\  {\bf B643}, 93 (2002);
D.~Cremades, L.~E.~Ib\'a\~nez and F.~Marchesano, JHEP 0207, 022 (2002);
J.~R.~Ellis, P.~Kanti and D.~V.~Nanopoulos,
  Nucl.\ Phys.\ B {\bf 647}, 235 (2002);
D.~Bailin, G.~V.~Kraniotis, and A.~Love, Phys.\ Lett.\ B {\bf 530}, 202
(2002); Phys.\ Lett.\ B {\bf 547}, 43 (2002); Phys.\ Lett.\ B {\bf 553}, 79
(2003); JHEP {\bf 0302}, 052 (2003); C.~Kokorelis, JHEP {\bf 0209}, 029
(2002); JHEP {\bf 0208}, 036 (2002);  JHEP {\bf 0211}, 027
(2002); Nucl.\ Phys.\ B {\bf 677}, 115 (2004); hep-th/0210200;
 hep-th/0406258;
 hep-th/0412035;
T.~Li and T.~Liu, Phys.\ Lett.\ B {\bf 573}, 193 (2003).

\bibitem{CSU1}
M.~Cveti\v c, G.~Shiu and A.~M.~Uranga, Phys.\ Rev.\ Lett.\  {\bf
87}, 201801 (2001).

\bibitem{CSU2}
M.~Cveti\v c, G.~Shiu and A.~M.~Uranga, Nucl.\ Phys.\ B {\bf 615},
3 (2001).

\bibitem{CP} M. Cveti\v c and I. Papadimitriou,
Phys.\ Rev.\ D {\bf 67}, 126006 (2003).

\bibitem{Cvetic:2002pj}
  M.~Cveti\v c, I.~Papadimitriou and G.~Shiu,
  Nucl.\ Phys.\ B {\bf 659}, 193 (2003)
  [Erratum-ibid.\ B {\bf 696}, 298 (2004)].

\bibitem{CLL}
M.~Cveti\v c, T.~Li and T.~Liu,
Nucl.\ Phys.\ B {\bf 698}, 163 (2004).

\bibitem{Cvetic:2004nk}
  M.~Cveti\v c, P.~Langacker, T.~Li and T.~Liu,
  Nucl.\ Phys.\ B {\bf 709}, 241 (2005).


\bibitem{Chen:2005ab}
  C.-M.~Chen, G.~V.~Kraniotis, V.~E.~Mayes, D.~V.~Nanopoulos and J.~W.~Walker,
  Phys.\ Lett.\ B {\bf 611}, 156 (2005); hep-th/0507232.


\bibitem{CLS1}
 M.~Cveti\v c, P.~Langacker and G.~Shiu,
Phys.\ Rev.\ D {\bf 66}, 066004 (2002).

\bibitem{CLS2}
M.~Cveti\v c, P.~Langacker and G.~Shiu, Nucl.\ Phys.\ B {\bf 642},
139 (2002).


\bibitem{CLW}
M.~Cveti\v c, P.~Langacker and J.~Wang,
Phys.\ Rev.\ D {\bf 68}, 046002 (2003).


\bibitem{ListSUSYOthers}
R.~Blumenhagen, L.~G\"orlich and T.~Ott, JHEP {\bf 0301}, 021 (2003);
G.~Honecker, Nucl.\ Phys.\  {\bf B666}, 175 (2003);
G.~Honecker and T.~Ott,
Phys.\ Rev.\ D {\bf 70}, 126010 (2004)
  [Erratum-ibid.\ D {\bf 71}, 069902 (2005)].


\bibitem{GVW}
S.~Gukov, C.~Vafa and E.~Witten,
Nucl.\ Phys.\ B {\bf 584}, 69 (2000) [Erratum-ibid.\ B {\bf 608}, 477 (2001)].


\bibitem{CU}
J. F. G. Cascales and A. M. Uranga,
JHEP {\bf 0305}, 011 (2003).

\bibitem{BLT}
R.~Blumenhagen, D.~L\"ust and T. R. Taylor,
Nucl.\ Phys.\ B {\bf 663}, 319 (2003).


\bibitem{Villadoro:2005cu}
  G.~Villadoro and F.~Zwirner,
  JHEP {\bf 0506}, 047 (2005).


\bibitem{Camara:2005dc}
  P.~G.~Camara, A.~Font and L.~E.~Ibanez,
  hep-th/0506066.


\bibitem{MS}
F.~Marchesano and G.~Shiu,
Phys.\ Rev.\ D {\bf 71}, 011701 (2005);
JHEP {\bf 0411}, 041 (2004).

\bibitem{CL}
M.~Cveti\v c and T.~Liu, Phys.\ Lett.\ B {\bf 610}, 122 (2005).


\bibitem{Kumar:2005hf}
  J.~Kumar and J.~D.~Wells,
  hep-th/0506252.


\bibitem{Cvetic:2005bn}
  M.~Cveti\v c, T.~Li and T.~Liu,
  Phys.\ Rev.\ D {\bf 71}, 106008 (2005).


\bibitem{CLLL-P}
  M.~Cveti\v c, P.~Langacker, T.~Li and T.~Liu, in preparation.


\bibitem{Soft BreakingI}
P.~G.~Camara, L.~E.~Ib\'a\~nez  and A.~M.~Uranga,
Nucl.\ Phys.\ B {\bf 689}, 195 (2004);
Nucl.\ Phys.\ B {\bf 708}, 268 (2005);
L.~E.~Ib\'a\~nez, Phys.\ Rev.\ D {\bf 71}, 055005 (2005);
A.~Font and L.~E.~Ib\'a\~nez , JHEP {\bf 0503}, 040 (2005).

\bibitem{Soft BreakingII}
D.~L\"ust, S.~Reffert and S.~Stieberger,
 hep-th/0410074; Nucl.\ Phys.\ B {\bf 706}, 3 (2005).


\bibitem{Soft BreakingIII}
F.~Marchesano, G.~Shiu and L.~T.~Wang, 
Nucl.\ Phys.\ B {\bf 712}, 20 (2005);
G.~L.~Kane, P.~Kumar, J.~D.~Lykken and T.~T.~Wang,
  Phys.\ Rev.\ D {\bf 71}, 115017 (2005).

\bibitem{CPY}
  M.~Cveti\v c and I.~Papadimitriou,
  Phys.\ Rev.\ D {\bf 68}, 046001 (2003)
  [Erratum-ibid.\ D {\bf 70}, 029903 (2004)].


\bibitem{LustetalIII}
  D.~L\"ust, P.~Mayr, R.~Richter and S.~Stieberger,
  Nucl.\ Phys.\ B {\bf 696}, 205 (2004).

\bibitem{Abel:2004ue}
  S.~A.~Abel and B.~W.~Schofield, JHEP {\bf 0506}, 072 (2005).

\bibitem{RGG} 
A. de R\'{u}jula, H. Georgi, and S. L. Glashow,
in \textit{Fifth Workshop on Grand Unification}, edited by K.
Kang, H. Fried, and P. Frampton (World Scientific, Singapore,
1984), p. 88.  For an earlier discussion, see Y. Achiman and B.
Stech, in \textit{New Phenomena in Lepton-Hardron Physics}, edited
by D. E. C. Fries and J. Wess (Plenum, New York, 1979), p. 303.

\bibitem{Babu:1985gi}
  K.~S.~Babu, X.~G.~He and S.~Pakvasa,
  Phys.\ Rev.\ D {\bf 33}, 763 (1986).


\bibitem{Choi:2003ag}
  K.~S.~Choi and J.~E.~Kim,
  Phys.\ Lett.\ B {\bf 567}, 87 (2003), and references therein.


\bibitem{Choi:2005pk}
  K.~S.~Choi and J.~E.~Kim,
  hep-th/0508149.


\bibitem{Giddings0105097}
  S.~B.~Giddings, S.~Kachru and J.~Polchinski,
  Phys.\ Rev.\ D {\bf 66}, 106006 (2002).

\bibitem{Kachru0201028}
  S.~Kachru, M.~B.~Schulz and S.~Trivedi,
  JHEP {\bf 0310}, 007 (2003).


\bibitem{Angelantonj:2005hs}
  C.~Angelantonj, M.~Cardella and N.~Irges,
  Nucl.\ Phys.\ B {\bf 725}, 115 (2005).


\bibitem{Witten9810188}
  E.~Witten,
  JHEP {\bf 9812}, 019 (1998).

\bibitem{Uranga0011048}
  A.~M.~Uranga,
  Nucl.\ Phys.\ B {\bf 598}, 225 (2001).


\end{thebibliography}
\end{document}